\documentclass[
    ,final            
  ]
  {aipproc}

\layoutstyle{8x11double}
\begin{document}
\title{Effect of a Quenched Disorder on the Order Parameter of Superfluid $^3$He}
\author{I. A. Fomin}{
  address={P. L. Kapitza Institute for Physical Problems,
ul. Kosygina 2, 119334 Moscow, Russia}}


\begin{abstract}
As a consequence of continuous degeneracy of the order parameter
of the superfluid $^3$He  quenched disorder in a form of aerogel
gives rise both to a disruption of the orientational long-range
order of the condensate and to a significant change of the order
parameter itself. There exist a class of quasi-isotropic  order
parameters which are not sensitive to the disorienting effect of
aerogel. While the BW order parameter belongs to this class the
ABM does not. Possible candidate for the order parameter of the
observed  A-like phase is discussed.
\end{abstract}

\maketitle

Average distance $\xi_a$ between the strands of high porosity
aerogels introduced as impurities in the superfluid $^3$He
\cite{parpia,halp1} does not differ much from the superfluid
coherence length $\xi_0$. This property is in a conflict with the
condition of applicability of the conventional theory of
superconducting alloys \cite{abri} $n\xi_0^3\gg 1$, where $n$ is
the average density of impurities. In aerogels $n\xi_0^3\sim 1$
and fluctuations of $n\xi_0^3$ are significant. In a vicinity of
the transition temperature $T_c$ effect of fluctuations can be
described phenomenologically by introduction of random terms in
the Ginzburg and Landau (GL) functional. For the scalar order
parameter $\psi$ such analysis has been made by Larkin and
Ovchinnikov \cite{LarkOv}. According to this analysis the most
singular at $T\to T_c$ corrections to the order parameter
originate from the term $F_{\eta}=N(0)\int \eta({\bf
r})|\psi|^2d^3r$ (here $N(0)$ is the density of states and
$\eta({\bf r})$ -- a random scalar), which describes local
variations of $T_c$. Generalization of this approach to the
3$\times$3 complex matrix order parameter $A_{\mu j}$ is
straightforward. Additional term in the GL functional has a form:
$$
F_{\eta}=N(0)\int \eta_{jl}({\bf r})A_{\mu j}A_{\mu l}^*d^3r .
\eqno(1)
$$
Now $\eta_{jl}({\bf r})$ is a random tensor with zero average
value. At the strength of $t\to -t$ invariance tensor
$\eta_{jl}({\bf r})$ is real and symmetrical, its isotropic part
$\eta^{(i)}_{jl}({\bf r})=\frac{1}{3}\eta_{mm}({\bf
r})\delta_{jl}$ describes  local variations of $T_c=T_c({\bf r})$
due to fluctuations of the density of scatterers, it couples only
to the overall amplitude of the order parameter $A_{\mu l}A_{\mu
l}^*\equiv\Delta^2$. Effect of $\eta^{(i)}_{jl}({\bf r})$ is
similar to that for the scalar case -- broadening of the
transition and suppression of the absolute value of the order
parameter. New properties originate from the anisotropic part
$\eta^{(a)}_{jl}({\bf r})\equiv\eta_{jl}({\bf
r})-\frac{1}{3}\eta_{ll}({\bf r})\delta_{jl}$ which describes
local splitting of  $T_c$ for different projections of angular
momenta because of breaking of spherical symmetry by the aerogel
strands. Contribution of $\eta^{(a)}_{jl}({\bf r})$ to $F_{\eta}$
depends on the particular form of the order parameter and on its
orientation with respect to the aerogel. Correspondingly
$\eta^{(a)}_{jl}({\bf r})$ can both influence the form of order
parameter and locally change its orientation. To make a
qualitative guess of possible effect of $\eta^{(a)}_{jl}({\bf r})$
for different order parameters one can start with a uniform order
parameter and substitute it into $F_{\eta}$.  No essential
difference with respect to the scalar case is expected if
$\eta^{(a)}_{j l}A_{\mu j}A_{\mu l}^*=0$. This is the case for the
BW-order parameter of bulk B-phase: $A^{BW}_{\mu j} = \Delta
e^{i\varphi}R_{\mu j}$, where $R_{\mu j}$ -- a real orthogonal
matrix, but not for the bulk A-phase (ABM). Its order parameter
has a form: $A^{ABM}_{\mu j}=(\Delta/\sqrt{2})\hat d_{\mu}(\hat
m_j+i\hat n_j)$ with $(\hat m_j,\hat n_j)=0$, then $
\eta^{(a)}_{jl}({\bf r})A_{\mu j}A_{\mu l}^*\sim
-\eta^{(a)}_{jl}l_jl_l\ne 0 $, (here $l_j=e_{jik}m_in_k$). This
property of ABM-order parameter is of importance in connection
with the identification of the A-like phase which has been
observed in liquid $^3$He in aerogel just below $T_c$
\cite{osher,dmit}.

 Disorienting effect of the random tensor
$\eta_{jl}({\bf r})$ on the ABM-order parameter has been
emphasized by Volovik \cite{volov1}. Following argument of Larkin
\cite{Larkin} and Imry and Ma \cite{imry} he concluded that there
may by no long-range order with the ABM-order parameter and
suggested instead a glass state. In that state order parameter
preserves ABM form locally but its orientation changes on a scale
$L_{IM}\sim \xi_0/\eta^2$. Characteristic value of $\eta^2$ can be
estimated e.g. from the measured width of the specific heat jump.
This width is determined by the parameter
$\alpha\equiv\eta^2/\sqrt{|\tau|}$ where $\tau=(T-T_c)/T_c$. A
region where fluctuations dominate is determined by the condition
$\alpha\sim 1$ \cite{LarkOv}. For $^3$He using the data of
Ref.\cite{he} for 22.5 bar one arrives at $\eta^2\sim 1/5$ and
$L_{IM}\sim 5\xi_0\sim 10^{-5}cm$.

Except for the disruption of the orientational order
$\eta^{(a)}_{jl}$ influences a form of the order parameter. For
the scalar case only amplitude of the order parameter can change
and its relative change is of the order of $\alpha\ll 1$. For the
triplet case the decrements of different components of $A_{\mu j}$
can be different.  Moreover their change can be not small in case
of degeneracy. For a triplet Cooper pairing there are altogether
18 extrema of the free energy \cite{march} and some of them can
come close in energy. Contribution of fluctuations to the energy
being small in comparison with the full condensation energy can be
comparable or larger then the small difference of energies of
competing states. As a result fluctuations can shift the
equilibrium and determine a choice of the order parameter. Let us
consider this possibility in more details.  With account of the
random term GL functional takes the form:
$$
 F_{GL}=N(0)\int d^3r[(\tau+\frac{1}{3}\eta_{ll}({\bf r})) A_{\mu j}A_{\mu j}^*+
$$
$$
 (\eta_{jl}({\bf r})- \frac{1}{3}\eta_{mm}({\bf
r})\delta_{jl})A_{\mu j}A_{\mu l}^* +
\frac{1}{2}\sum_{s=1}^5\beta_sI_s+
$$
$$
\frac{1}{2}\left(K_1\frac{\partial A_{\mu l}}{\partial x_j}
\frac{\partial A^*_{\mu l}}{\partial x_j}+ K_2\frac{\partial
A_{\mu l}}{\partial x_j}\frac{\partial A^*_{\mu j}}{\partial
x_l}+K_3\frac{\partial A_{\mu j}}{\partial x_j}\frac{\partial
A^*_{\mu l}}{\partial x_l}\right)\Bigr] ,           \eqno(2)
$$
where  $I_s$ -  4-th order invariants in the expansion of the free
energy over $A_{\mu j}$. Coefficients $\beta_1,...\beta_5,
K_1,K_2,K_3$ -- phenomenological constants.

Because of a presence of {\bf r}-dependent tensor $\eta_{jl}({\bf
r})$ in the functional (2) solutions of the corresponding
GL-equations also depend on {\bf r}. It is convenient to separate
the order parameter into its average $\bar A_{\mu j}$ and
fluctuating part $a_{\mu j}({\bf r})$ so that $A_{\mu j}({\bf
r})=\bar A_{\mu j}+a_{\mu j}({\bf r})$. Below $T_c$ \quad $\bar
A_{\mu j}\ne 0$. The terms proportional to $\eta_{jl}({\bf r})$ in
Eq. (2) dominate at $T\to T_c$. Within a region $|\tau|\leq\eta^4$
or $\alpha\geq 1$ \quad $a_{\mu j}({\bf r})$ gives at least as
important contribution to the physical quantities as $\bar A_{\mu
j}$.  It is difficult to analyze solutions of GL-equations within
this region.  Analysis is possible when $\alpha\ll 1$ and $|a_{\mu
j}|\ll |\bar A_{\mu j}|$. In this region of parameters the
iteration procedure of Ref.\cite{LarkOv} renders equations for
$\bar A_{\mu j}$:
$$
\tau\bar A_{\mu j}+\frac{1}{2}\sum_{s=1}^5
\beta_s\bigl[\frac{\partial I_s}{\partial A^*_{\mu j}}+
\frac{1}{2}\bigl( \frac{\partial^3 I_s}{\partial A^*_{\mu
j}\partial A_{\nu n}\partial A_{\beta l}} <a_{\nu n}a_{\beta l}>+
$$
$$ 2\frac{\partial^3
I_s}{\partial A^*_{\mu j}\partial A^*_{\nu n} \partial A_{\beta
l}}<a^*_{\nu n}a_{\beta l}>\bigr)\bigr]= -<a_{\mu l}\eta_{lj}>.
\eqno(3)
$$

\smallskip
and $a_{\mu j}({\bf r})$:
$$
\tau a_{\mu j}+\frac{1}{2}\sum_{s=1}^5\beta_s\bigl[
\frac{\partial^2 I_s}{\partial A^*_{\mu j}\partial A_{\nu
n}}a_{\nu n}+ \frac{\partial^2 I_s}{\partial A^*_{\mu j}\partial
A^*_{\nu n}}a^*_{\nu n}\bigr]-
$$
$$ \frac{1}{2}K\left(\frac{\partial^2
a_{\mu j}}{\partial x_l^2}+ 2\frac{\partial^2 a_{\mu l}}{\partial
x_l \partial x_j}\right)= -\bar A_{\mu l}\eta_{lj},  \eqno(4)
$$

 Equations (3),(4) have the same structure as in the scalar
case. Linear equation for $a_{\mu j}$ is solved by  Fourier
transformation and its solution has the following structure:
$$
a_{\mu j}({\bf k})\sim\frac{A_{\mu l}\eta_{lj}({\bf
k})}{(\kappa^2+k^2)}, \quad \kappa\sim 1/\xi. \eqno(5)
$$
Then for the average binary products entering Eq. (3) we have
$$
<a_{\mu j}a_{\nu n}>\sim A_{\mu l}A_{\nu
m}\int\frac{<\eta_{lj}({\bf k})\eta_{mn}({\bf
-k})>}{(\kappa^2+k^2)^2}\frac{d^3k}{(2\pi)^3}. \eqno(6)
$$
These terms in Eq. (3) give corrections of a relative order of
$\alpha\ll 1$ except for directions in which $\kappa=0$ (Goldstone
directions) when the integral diverges. This is the case for the
increments of $\bar A_{\mu j}$ and $\bar A^*_{\mu j}$ at the
infinitesimal rotation $\theta_n$:
$$
\omega_{\mu j}=\theta_ne^{jnr}\bar A_{\mu r}, \omega^*_{\mu
j}=\theta_ne^{jnr}\bar A^*_{\mu r}.         \eqno(7)
$$

The divergency means that contribution of the averaged products of
projections of
 $a_{\beta l}$ in directions determined by $\omega_{\mu j}$ to Eq. (3) is of
a relative order of $\alpha^{\lambda}$,\quad $\lambda<1$.  The
value of $\lambda$ depends on a mechanism of cutting-off the
divergency. The mechanism is provided by anharmonicity. Previously
suggested evaluation of the diverging integral from the condition
$|a_{\beta l}|\sim |\bar A_{\beta l}|$ \cite{repl1} overestimates
it and renders $\lambda=0$. Better estimation is obtained by
keeping anharmonic (third order) terms in Eq. (4). The obtained
nonlinear equation can be treated within a mean-field approach,
i.e. the third-order terms are substituted as $aaa \to a<aa>$,
where $a$ stays for any of components of $a_{\mu j}$  and only
diverging binary averages $<aa>_{sing}$ are kept. Thus obtained
linear equation has solution of a symbolic form (all indices are
suppressed):
$$
a\sim\frac{Q\eta}{\Delta[(\xi_0k)^2+\beta <aa>]}, \eqno(8)
$$
where $\beta\sim 1/T_c^2$ is a typical value of
$\beta$-coefficients in the GL-expansion and $Q$ is a tensor
$Q_{rl}=\bar A_{\mu r}\bar A^*_{\mu l}+\bar A_{\mu l}\bar A^*_{\mu
r}$. For binary averages in the Goldstone directions we have
$$
<aa>_{sing}=const.\frac{Q^2}{\Delta^2}\int\frac{<\eta\eta>k^2dk}{[(\xi_0
k)^2+\beta<aa>_{sing}]^2}.    \eqno(9)
$$
This is an equation for $<aa>_{sing}$. Its solution has a form
$$
<aa>_{sing}^{3/2}=const.\frac{Q^2<\eta\eta>_0}{\Delta^2\sqrt{\beta}\xi_0^3}.
\eqno(10)
$$
Here $<\eta\eta>_0=\int<\eta(0)\eta({\bf r})>d^3r$. With the aid
of Eq.(8) $<aa>_{sing}$ can be estimated as
$<aa>_{sing}\sim(\Delta Q^2\alpha)^{2/3}$. At $\alpha\to 0$ the
singular terms in the GL-equation give a principal part of the
contribution of fluctuations to this equation.
 GL-equation now has a structure:
$$
\frac{\delta}{\delta A_{\mu
j}^*}[F^{(0)}+F^{fl}_{reg}+F^{fl}_{sing}]=0. \eqno(11)
$$
Consider a set of competing states which are nearly degenerate
with ABM:
$$
\left|\frac{F_c-F_{ABM}}{F_{ABM}}\right|\equiv\varepsilon\ll 1 ,
\eqno(12)
$$
then, in  Eq. (11) the first term is of the order of
$\varepsilon$, the second - of the order of $\alpha$ and the third
-- of the order of $(\alpha)^{2/3}$. A character of solution
depends on a relative values of small parameters $\varepsilon$ and
$(\alpha)^{2/3}$. If $\varepsilon\gg(\alpha)^{2/3}$ the order
parameter is determined as a minimum of the unperturbed
GL-functional $F^{(0)}$ (supposedly ABM-state) and fluctuations
give small corrections to its form. In the opposite limit
$\varepsilon\ll(\alpha)^{2/3}$ the form of the order parameter is
determined by the fluctuations:
$$
\frac{\delta}{\delta A_{\mu j}^*}[F^{fl}_{sing}]=0. \eqno(13)
$$
This equation is satisfied if $Q_{jl}\sim\delta_{jl}$:
$$
A_{\mu j}A_{\mu l}^*+A_{\mu l}A_{\mu j}^*=const.\delta_{jl},
\eqno(14)
$$
or, equivalently:
$$
\eta^{(a)}_{jl}A_{\mu j}A_{\mu l}^*=0.  \eqno(15)
$$
As has been discussed above, for the order parameters, meeting
condition (15) there is no disorienting effect of the random
tensor $\eta^{(a)}_{jl}$ and long-range order is preserved.
 Generally this condition specifies a degenerate family of
so-called "robust", or quasi-isotropic  states. For partial
lifting of the degeneracy the remaining terms in the free energy
have to be minimized
$$
\frac{\delta}{\delta A_{\mu j}^*}[F^{(0)}+F^{fl}_{reg}]=0
\eqno(16)
$$
with Eq. (14) as a constraint. Eq.(16) determines in particular
the overall amplitude $\Delta$. It has to be verified then that
the found solutions satisfy condition (12). If it is the case
there exist a temperature region where  condition
$\varepsilon\ll(\alpha)^{2/3}$ is met as well since $\varepsilon$
does not depend on temperature but $\alpha$ grows as
$1/\sqrt{|\tau|}$ at $T\to T_c$. Within this temperature region
the above procedure of finding solution of GL-equation which are
determined by fluctuations is justified.

In case of A-like phase the competing states have to be of equal
spin pairing (ESP) type. It follows from the experimental
observation that magnetic susceptibility of the emerging phase
coincides with that of normal and ABM-phases \cite{osher}. A
general form of the order parameter for ESP-state is:
$$
A^{ESP}_{\mu j}=\Delta\frac{1}{\sqrt{3}}[\hat d_{\mu}( m_j+i n_j)+
\hat e_{\mu}( l_j+i p_j)], \eqno(17)
$$
where $\hat d_{\mu}$ and $\hat e_{\mu}$ are mutually orthogonal
unit vectors, $m_j,n_j,l_j,p_j$ -- arbitrary real vectors. The
order parameter $A^{ESP}_{\mu j}$ meets criterion  (14)
 if vectors $m_j,n_j,l_j,p_j$ satisfy the equation
$$
m_jm_l+n_jn_l+l_jl_l+p_jp_l=\delta_{jl}. \eqno(18)
$$
A detailed description of this solution is presented elsewhere
\cite{fom2}. If additional constraint of the absence of
spontaneous magnetization is imposed the obtained solution can be
rewritten as:
$$
A^R_{\mu j}=\Delta\frac{1}{\sqrt{3}}e^{i\psi}[\hat d_{\mu}(\hat
b_j+i\cos\gamma\hat c_j)+ \hat e_{\mu}(\hat a_j+i\sin\gamma\hat
c_j)], \eqno(19)
$$
where $\hat a_j,\hat b_j,\hat c_j$ -- mutually orthogonal unit
vectors, $\psi$ and $\gamma$ -- parameters.

Now the condition (12) has to be verified. For the order parameter
Eq. (19) $\varepsilon$ is expressed in terms of coefficients
$\beta_i$ as:
$$
\varepsilon=\frac{\beta_{13}-4\beta_{45}}{9\beta_2+\beta_{13}+5\beta_{45}},
\eqno(20)
$$
where $\beta_{13}=\beta_1+\beta_3$ etc. For the weak coupling
values of $\beta$-coefficients  $\varepsilon=1/19$ i.e. small,
but in that case BW-phase is more favorable then ABM. The most
important are strong coupling corrections to the combination
$\beta_{45}$ since
 the weak coupling value $\beta_{45}=0$ determines a boundary between ABM and
axiplanar phases in the space of $\beta$-parameters \cite{mermin}.
For the moment it is difficult to make definite conclusions about
the value of $\beta_{45}$ realized in liquid $^3$He. Combinations
of $\beta_1,...\beta_5$ deduced from the analysis of thermodynamic
data \cite{gould} do not form a complete set and can not define
$\varepsilon$ without ambiguity. On the other hand the values of
$\beta_1,...\beta_5$ found theoretically from model assumptions
\cite{sauls} do not agree with the experimental findings in
particular about $\beta_5$.

Eq. (19) gives solution of GL-equation of a zeroth order on the
small parameter $\varepsilon/\alpha^{2/3}$. In the real $^3$He
this ratio can be not very small so that the finite deviation from
 "robust" state occur and disorienting effect of the random
tensor $\eta_{jl}({\bf r})$ comes into play. The outlined
procedure offers in that case only a local form of the order
parameter. Principal contribution to the integral Eq.(9) comes
from the wave vectors $(\xi_0k)^2\sim \beta<aa>_{sing}$, or length
scales $L_{c-off}\sim\xi/(\alpha q^2)^{1/3}$, here
$q^2=Q^2/\Delta^2$ is the normalized deviation of the order
parameter from the "robust" form. The local form of the order
parameter is preserved if it changes its orientation in space on a
length scale which is much greater then $L_{c-off}$. This is the
case for Imry and Ma disorienting effect: $L_{IM}/L_{c-off}\sim
1/(\alpha q^2)^{2/3}\gg 1$. The resulting glass-type state is
based on  a "nearly robust" state and the characteristic size of
the "domains" is much greater then in the case of the ABM-based
glass state. The most important difference between the two states
is that the order parameter $A^R_{\mu j}$  unlike the
$A^{ABM}_{\mu j}$ does not have the combined gauge-orbital
symmetry. As a consequence the property of superfluidity for
$A^R_{\mu j}$ is preserved even in the glass-type state. Formally
it means that the averages of a type $<A^R_{\mu j}({\bf
r})A^R_{\nu l}({\bf r})>$ are finite and can be used as the order
parameter of the "superfluid glass" state \cite{vol-khm}.

So, one can conclude that at a triplet Cooper pairing quenched
disorder of a type of "fluctuations of the transition temperature"
$\eta_{jl}({\bf r})$  influences both  form of the order parameter
and its local orientation. There are "robust" classes of order
parameters, for which disorienting effect disappears. If the form
of the order parameter required by the minimization of the initial
free energy is not "robust" there is a mechanism of adjustment
which tends to bring the order parameter closer to the "robust"
form and to decrease the disorienting effect of the random tensor
$\eta_{jl}({\bf r})$. Both disorienting effect and the mechanism
of adjustment originate from degeneracy of the unperturbed free
energy with respect to orbital rotations of the order parameter.

I acknowledge the hospitality of the Laboratory of Atomic and
Solid State Physics of Cornell University, where final part of
this work has been done.  This research was supported in part by
RFBR grant (no. 04-02-16417), by Ministry of Education and Science
of Russian Federation and by CRDF grant RUP1-2632-MO04.

  \end{document}